\documentclass{article}
\usepackage[utf8]{inputenc}
\usepackage{authblk}
\usepackage{array}
\title{\emph{iid2022}: A Workshop on Statistical Methods for Event Data in Astronomy} 

\author[1]{Eric D. Feigelson}
\author[2]{Massimiliano Bonamente}
\affil[1] {Center for Astrostatistics and Department of Astronomy \& Astrophysics, Penn State University, University Park PA 16802 ~~ e5f@psu.edu}
\affil[2]{Department of Physics and Astronomy, University of Alabama,
 Huntsville AL 35899 ~~ max.bonamente@uah.edu}

\begin{document}

\maketitle

\begin{center}
Appears in Frontiers in Astronomy and Space Sciences,  2023 \\ DOI 10.3389/fspas.2023.1228508
\end{center}

\begin{abstract}
    We review the \emph{iid2022} workshop on statistical methods for X-ray and $\gamma$-ray astronomy and high--energy astrophysics event data in astronomy, held in Guntersville, AL, on Nov. 15-18 2022.  New methods for faint source detection, spatial point processes, variability and spectral analysis, and machine learning are discussed.  Ideas for future developments of advanced methodology are shared. 
\end{abstract}

\section{Statistical Challenges Arising in High--Energy Astrophysics}

The science analysis of data in high--energy astrophysics differs from most fields of astronomy in important ways.  The data, typically from space-based observatories, consist of energetic photons counted individually as they arrive in a detector.  These datasets often can be viewed in tabular form as a sequence of events with four characteristics: arrival time, location in two-dimensions, and energy.  The analysis commonly proceeds in stages: sources are identified in the 2-dimensional image, photons are extracted for individual sources or emitting regions, and 1-dimensional analysis proceeds for the energy distribution and arrival times.  These univariate distributions are often complicated: multi-component spectral emission processes are convolved with instrumental sensitivity, and temporal processes can depend on unpredictable variations in accretion onto compact objects.  Common analysis procedures include:   
\begin{enumerate}
\item Individual photons are examined, often smoothed with knowledge of the telescope point spread function, in the image plane;
\item Sparse samples of individual events from faint sources are modeled along one-dimensional energy (spectra) or temporal axis (light curves);
\item Richer samples of events are grouped into bins along the spectral or temporal axis and then subject to statistical or astrophysical modeling.  
\end{enumerate}

Table 1 summarizes important statistical procedures developed in the high--energy astrophysical community over the past half century.  The accomplishments are impressive, but the impact on the research community is mixed.  Some methods, such as the Lomb-Scargle periodogram, are widely used, although there may be insufficient appreciation of the challenges of estimating reliable False Alarm Probabilities \cite{VanderPlas18}.  But other valuable statistical procedures $-$ such as different limits for source existence and flux \cite{Kashyap10} and Bayesian estimates of faint-source hardness ratios \cite{Park06} $-$ are not commonly used. Many have listened the warning that likelihood ratio tests should not be used near the boundary of parameter values \cite{Protassov02}, but there is inadequate recognition that likelihood ratios should be penalized by model complexity as with the Bayesian Information Criterion. 

There is also a general unawareness within the astronomical community of basic methods that are common in other fields. For example, multiple linear regression for count data \cite{cameron2013} is used extensively in econometrics and other areas, but astronomers often compare a response variable to single covariates in a sequential fashion. Aperiodic stochastic temporal behaviors (that might arise from accretion processes or magnetic activity) are analyzed using Fourier methods designed for periodic time series rather than autoregressive modeling \cite{Box15}.

\section{The \emph{iid2022} Workshop}

These issues motivated the workshop \emph{iid2022: Statistical Methods for Event Data Illuminating the Dynamic Universe}  workshop, held in Huntsville Alabama on November 15-18, 2022. The spirit of the workshop was
to give the participant an opportunity to review and learn about certain statistical methods, and also make presentations based on their own research. Accordingly, the eight sessions had introductory talks by more senior scientists, followed by oral presentations by students and early--career scientists. The \emph{National Science
Foundation} provided support for twenty students and early--career scientists to attend the workshop, via a grant issued to the University of Alabama in Huntsville.
Such support was essential to attract students who would not otherwise have had the opportunity to attend. 

Table~1 lists presentations made at the workshop.
The vast majority of attendees were astronomers, with a few notable exceptions such as Prof. Dale Zimmermann of the University of Iowa, who gave the keynote lecture, and biostatistics graduate student Jesus Vasquez from the University of North Carolina at Chapel Hill.

\vspace*{0.0in}

\begin{center}
{\large\bf Table 1: Presentations at the \em{iid2022} Workshop} \\  ~\\
\begin{tabular}{| p{2.8in}p{1.6in} |} \hline
~ & \\
\multicolumn{2}{|c|}{\bf KEYNOTE LECTURE} \\
Parametric estimation of spatial point ~~~~~~~~~~ processes & D. Zimmermann (Iowa) \\ 
~ & ~ \\
\multicolumn{2}{|c|}{\bf 1. STATISTICAL MODELING OF COUNT DATA} \\
Overview of regression methods for count data & M. Bonamente (UAH)\\
Bayesian field-based likelihood analysis & A. Heavens (Imperial) \\
A ground-based search for terrestrial gamma-ray flashes in Fermi-GBM Data & S. Lesage (student) \\ 
~ & ~ \\
\multicolumn{2}{|c|}{\bf 2. STATISTICS FOR IDENTIFICATION OF LOW-COUNT SOURCES} \\
Flux estimation from count-based data & D. Mortlock (Imperial)\\
Introduction to low-count statistics & V. Kashyap (Smithsonian)\\
Maximum likelihood calibration of the tip of the red  giant branch using Milky Way field giants & Siyang Li (student) \\
Are giant planet-hosting stars young? & C. Swastik (student)\\ 
~ & ~ \\

\multicolumn{2}{|c|}{\bf 3. ANALYSIS OF SPECTRAL DATA} \\
Goodness-of-fit for regression with count data & M. Bonamente (UAH) \\
Joint spatio-spectro-temporal analysis of X-ray events & V. Kashyap (Smithsonian)\\
Machine learning to detect CIV absorption lines in SDSS spectra & R. Monadi (student) \\
Properties of late-type dwarfs using low-resolution  spectroscopy from Gaia & Z. Way (student) \\
Assessing the impact of narrow-band information in photometric surveys & L. Nakazono (student)\\
Properties of lowest metallicity galaxies at z = 0.2 - 1 & 
I. Laseter (student) \\ 
~ & ~ \\
\multicolumn{2}{|c|}{\bf 4. STATISTICS FOR VARIABLE SOURCES}  \\
Time domain astronomy: Grouping to reveal structure & G. Belanger (ESAC) \\
Bayesian Basics & T. En{\ss}lin (MPA)  \\
Statistics for low count rate variable sources & E. Feigelson (Penn State)\\
Bayesian Blocks I & J. Scargle (NASA-Ames)\\
Bayesian Blocks II & M. Kerr (SLAC)\\
~ & ~ \\
\multicolumn{2}{|c|} {\bf 5. STRUCTURES IN IMAGES: SPATIAL POINT PROCESSES}  \\
Spatial Point Processes & E. Feigelson (Penn State)\\
Information field theory for event data & T. En{\ss}lin (MPA) \\ \hline
 \end{tabular}

\newpage 
 \begin{tabular}{| p{2.8in}p{1.6in} |} \hline
~ & ~ \\
Constraints on the temperature-density relation of the ISM  with non-negligible absorber spatial structure & T. Ksenia (student) \\
Image deconvolution and reconstruction methods in Poisson images & A. Siemiginowska (Smithsonian) \\
pyLira tutorial & A. Donath (student)\\
Phase coherence of the solar wind turbulence & M. Nakanotani (student) \\
Galaxy Clusters and Protoclusters at HST-SSP Survey & V. Marcelo (student)\\
Sampling Methods & T. Ensslin (MPA) \\
\multicolumn{2}{|c|}{\bf 6. APPLICATIONS TO ASTRONOMICAL DATA }  \\
Chemodynamical ages of small-scale kinematic structures in the solar neighborhood & I. Medan (student) \\
Modeling quasar UV/optical variability as stochastic diffusion processes & Weixiang Yu  (student)\\
Forming supermassive black holes with the collapse of Self-Interacting Dark Matter Halos & 
S. Gad-Nasr (student) \\
The Galactic Center as a gravitational laboratory & R. Della Monica (student) \\
A new method of investigation of the orientation of galaxies in clusters in the absence of information on their morphological types & M. Błażej (student) \\
Practical application of the new method of investigation of the alignment of galaxies in clusters &  W. Godłowski (Opolski) \\
~ & ~ \\
\multicolumn{2}{| c |}{\bf 7. NON-DETECTION: CENSORED AND TRUNCATED DATA} \\
Non-detections: Censoring and truncation in astronomical surveys & E. Feigelson (Penn State) \\
Using biostatistics doubly robust estimators to model the progression of symptoms for neurodegenerative diseases in astronomy & J. Vazquez (student) \\
~ & ~ \\
\multicolumn{2}{|c|}{\bf 8. MACHINE LEARNING AND NUMERICAL METHODS}  \\
From histograms to hierarchical models: Bayesian modeling of event and population data with point processes  & T. Loredo (Cornell)\\
Cosmology in the machine learning era & F. Villaescusa-Navarro \\
~ & (Simons Fnd) \\
Numerical information field theory & P. Franck (MPA) \\
Techniques for variational inference &  P. Franck (MPA)  \\
Estimating the sensitivity of a polarimeter module for the Large Area burst Polarimeter (LEAP) &  K. Oñate Melecio  (student) \\
\hline
\end{tabular}
\end{center}

\newpage 
\begin{center}
 \begin{tabular}{| p{2.8in}p{1.6in} |} \hline
~ & ~ \\
Machine learning applied to meteor detection filtering & S. Anghel (student)\\
Utilizing a Metropolis-Hastings algorithm to determine the dominant mechanism of particle energy gain by interacting with dynamic small-scale flux ropes & K. Van Eck (student)\\
A global 21cm signal emulator of 21cmFAST & D. Breitman (student)\\
 & \\
Concluding remarks & M. Bonamente (UAH) \\
\hline
\end{tabular}
\end{center}
\vspace*{0.in}

\section{Past Accomplishments in Methodology }

High--energy astronomy has its roots in the study of cosmic rays on mountaintops during the 1930s and the discovery of X-rays from the solar corona during the 1950s \cite{Rossi48, Tousey51}.  The first detection of X-rays outside the Solar System involved a few thousand counts from the Galactic Plane obtained during a brief rocket flight \cite{Giacconi1962}.  Early analyses involved simple statistical procedures such as the running mean \cite{Bowyer64} or (mathematically incorrect) least squares procedures applied to Poisson distributed data. The first use of the Poisson distribution to derive a cosmic source flux upper limit appears to be by Hearn (1968).

As satellite observatories replaced sounding rockets, more specialized statistical procedures began to emerge and accelerated in the early 21st century.  Table 2 lists some of the important milestones classified by the scientific problem addressed. Some methods have had very broad impact with over a thousand citations by later studies. Altogether, the development and promulgation of analysis methods has been substantial and often quite successful.  

In addition to procedures developed by practitioners within the field, methods for astronomy have been adopted from the wider arena of statistics. In early years, the textbook \emph{Data Reduction and Error Analysis for the Physical Sciences} \cite{bevington1969} promoting least squares procedures had the greatest impact, not least because it included convenient Fortran codes that could be typed into IBM cards and used on main frame computers.  It was largely supplanted by {\it Numerical Recipes: The Art of Scientific Computing} (Press et al. 1992) with editions providing code in Fortran, Pascal, C and C++.  Numerical Recipes garnered  $>$12,000 citations in astronomy and $>$120,000 citations in all fields. 

Other useful textbooks include \emph{Statistical Methods in Experimental Physics} \cite{eadie1971}, \emph{Practical Statistics for Astronomers}  \cite{wall2012}, \emph{Modern Statistical Methods for Astronomy with R Applications} [Feigelson \& Babu 2012], \emph{Statistics, Data Mining, and Machine Learning in Astronomy} [Ivezić et al. 2019],  and \emph{Statistics and Analysis of Scientific Data} \cite{bonamente2022book}.  Bayesian inference has become an important tool for modeling astronomical data as treated in texts like \cite{hilbe2017} and [Bailer-Jones 2017]. However, neither the classic works nor the newer volumes emphasize low-count rate problems as encountered in high--energy astronomy.   Some require a basic knowledge of probability and statistics, and this can  limit their diffusion among astronomers who are often missing such courses in their undergraduate education.  
\\ 

\begin{center}
{\large\bf Table 2: Statistical Milestones for \\ X-ray and $\gamma$-ray Astronomy} \\ ~\\ 
\begin{tabular}{| lrl |} \hline
{\bf Procedure}  & {\bf Citations} & {\bf Reference}\\

Likelihood-based model fitting& 2200~~~ & \cite{Cash79}  \\
                              & 1000~~~ & \cite{Mattox96} \\
                              & 600~~~ & \cite{Akritas1996} \\
                              & 500~~~ & \cite{Protassov02} \\

Faint source significance     & 1000~~~ &  \cite{Li83} \\
                              & 400~~~ & \cite{Kraft91} \\
                              & 50~~~ & \cite{Kashyap10} \\

Treatments of upper limits    & 300~~~ & \cite{Schmitt85} \\
                              & 700~~~ & \cite{Feigelson85} \\ 
                              & 700~~~ & \cite{Isobe86} \\
Truncation and selection effects  & 400~~~ & \cite{Mantz2010}\\
                                 &  300~~~ & \cite{Mantz2010b}\\
Searching for periodicity &  5000~~~ & \cite{Scargle82} \\
                        & 150~~~ & \cite{Leahy83} \\
                        & 300~~~ & \cite{deJager89} \\
                        & 230~~~ & \cite{Vaughan05} \\

Faint source detection & 190~~~ & \cite{Damiani97} \\
                           & 500~~~  & \cite{Freeman02} \\
                           & 130~~~ & \cite{Ebeling06}\\
                           & 180~~~ & \cite{Diehl06} \\
                           &  80~~~ & \cite{Zhang08} \\
                           & 240~~~ & \cite{Broos10} \\
                           & 10~~~ & \cite{Stein15} \\

Variability characterization & 1100~~~ & \cite{Edelson88} \\
                           & 260~~~ & \cite{Scargle98} \\
                           & 800~~~ & \cite{Vaughan03} \\
                           & 300~~~ & \cite{Uttley05} \\
                           
Hardness ratio             & 250~~~ &  \cite{Park06} \\

Treatments of measurement error & 900~~~ & \cite{Kelly07} \\
                                & 1,900~~~ & \cite{Gehrels1986}\\

Bayesian spectral modeling  &  70~~~ & \cite{vanDyk01} \\
                            & 800~~~ & \cite{Buchner14} \\
                            & 20~~~ & \cite{Xu14} \\
Markov Chain Monte Carlo    & 40~~~ & \cite{bonamente2004}\\
                            & 300~~~ & \cite{bonamente2006}\\
\hline
\end{tabular} 
\end{center}
\newpage

\begin{center}
{\large\bf Table 3: Some Statistical Methodology Featured \\ at the {\it iid2022} Workshop} \\  ~ \\

\begin{tabular}{| m{3.0in}| m{1.4in} | } \hline

Spatial Point Patterns: Methodology and Applications in R & \cite{Baddeley15} \\ ~& \\

A semi-analytical solution to the maximum-likelihood fit of Poisson data to a linear model using the Cash statistic & [Bonamente \& Spence 2022] \\ ~& \\

 LIRA {\textemdash} The Low-Counts Image Restoration and Analysis Package: A Teaching Version via R & \cite{Connors11} \\ ~& \\
 
Time domain methods for X-ray and gamma-ray astronomy & \cite{Feigelson22} \\ ~& \\

Matching Bayesian and frequentist coverage probabilities when using an approximate data covariance matrix & \cite{Percival22} \\ ~& \\

The denoised, deconvolved, and decomposed Fermi $\gamma$-ray sky. An application of the D$^{3}$PO algorithm & \cite{Selig15} \\ ~& \\

Studies in Astronomical Time Series Analysis. VI. Bayesian Block Representations & \cite{Scargle13} \\ ~& \\

Change-point Detection and Image Segmentation for Time Series of Astrophysical Images & \cite{Xu21} \\

\hline
\end{tabular}
\end{center}
\vspace*{0.3in}

Table 3 lists a few of the methods discussed in the {\it iid2022} workshop that are directly relevant to high--energy data and science analysis.  Software implementation are combined with methodologies to allow quick implementation.   In some cases, such as Baddeley's book for analyzing Poisson images and variability detection procedures discussed by Feigelson, the codes are already available in the general purpose R statistical software environment.  In other cases, such as Scargle's Bayesian Blocks and Xu's multidimensional change-point analysis, codes are written specifically for use in X-ray and $\gamma$-ray astronomy.

\section{Looking Towards the Future}

Presentations at the {\it iid2022} workshop demonstrate that the development of innovative procedures for analyzing high--energy astronomical data is proceeding in a vibrant fashion.  But there are considerable difficulties in promulgation of new methodology in the research communities.  We outline here challenges that can be readily identified and suggest directions for improvements for the coming years.  

{\bf Statistics Education} ~~ One of the main needs in high--energy astronomy is a more rounded background in statistics for its practitioners. Most graduate degrees leading to an advanced degree in astronomy or astrophysics have no requirement of statistics courses, and are often limited to a course on `data analysis methods' that lacks a foundation on statistical principles.  Astronomers should be familiar with differences between nonparametric hypothesis testing and parametric modeling, Poisson and Gaussian distributions, least squares and likelihood based modeling, and stationary and nonstationary processes.  Wavelet transforms, local regression, autoregressive models, and Fourier approaches to time series analysis should be taught.  

As both authors and teachers, it is our opinion that the typical high--energy data analyst should have a background that includes at least one undergraduate course using a statistics textbook such as \emph{Probability and Statistical Inference} \cite{hogg2023}. Such background would be beneficial to understand in detail the main statistical methods available, while giving the basic tools to undertake more complex tasks such as developing new statistical
methods.  At the graduate level, a course in methodology using textbooks like {\em Statistics, Data Mining, and Machine Learning in Astronomy: A Practical Python Guide for the Analysis of Survey Data} \cite{ivezic2019} and {\em Modern Statistical Methods for Astronomy with R Applications} [Feigelson and Babu 2012] should be widely available in astronomy departments.  

{\bf Integrate statistics into high--energy mission projects} ~~ 
High--energy astrophysics missions have traditionally included costs for `software development' to write pipelines for processing telemetry data through Level 1 and Level 2 data products.  But it is also important to fund, at the early stages, study of methods to be implemented in the pipeline and off-line science analysis by individual scientists.  Methods as simple as maximum--likelihood analysis of count data \cite{Cash79} and as complex as information theory for gamma-ray astronomy \cite{Ensslin2019} and 4-dimensional change-point analysis \cite{Xu21} should be considered. 

Centralized facilities like NASA's High Energy Astrophysics Science Archive Center and ESA's European Space Astronomy Centre should institute organized procedures to evaluate newer methodologies and bring them into their code libraries for use by the research communities.  Some methods can be incorporated into important existing software tools such as  \emph{XSPEC} \cite{arnaud1996}
and \emph{SPEX} \cite{kaastra1996}, while other methods would be stand-alone codes added to libraries such as \emph{HEASoft}.  Documentation and tutorials for training community scientists in methodology should accompany software releases.  

{\bf Funding for methodology} ~~  For two decades starting in 1990, NASA's Science Mission Directorate had an Applied Information Systems Research program that included development of statistical tools, machine learning procedures, computational methods and algorithms for astronomical missions.  But this program has changed focus and there is now no avenue for the research community to obtain funds for the development of new methodology for high--energy astrophysics.  A program is needed similar to NASA's Earth Science Division's Advanced Information Systems Technology Program that includes development of advanced tools for data and science analysis.  Several White Papers were submitted to the National Academy of Science Astro2020 Decadal Survey arguing for improved funding in astrostatistics and astroinformatics for all branches of the field. 

{\bf Attitudes towards advances in methodology} ~~ A major reason for the slow advancement in usage of advanced $-$ or even statistically acceptable $-$ statistical methods in high--energy astrophysics is absence of penalty for inaccurate or misleading analysis methods.  This includes review during mission planning, individual observing proposals, and the final published astrophysical literature.  Sometimes forces lean towards mundane analysis procedures: authors who present advanced statistical methods in an astrophysics paper might encounter a reviewer poorly prepared in statistics. The journals of the American Astronomical Society now have a Statistics Editor, and reviewers expert in statistical analysis can be sought in addition to a reviewer expert in the scientific topic.  A two-reviewer process is common for journals like \emph{Annals of Applied Statistics} and \emph{Journal of Applied Statistics}.  The high--energy research community that widely encourages improvements in telescope and detector capabilities should also encourage improvements in data analysis capabilities that can improve the scientific return from any instrument or observing project.   

\vspace*{0.3in}

\section*{Acknowledgments} MB gratefully acknowledges University of Alabama in Huntsville students Stephen Lesage, Juan Alonso Guzm\'{a}n and Samuel Johnson, whose dedication and support was essential for the success of the workshop. The \emph{iid2022} workshop was supported by NSF grant 2223560 `\emph{Conference: iid2022: Statistical Methods for Event Data - Illuminating the Dynamic Universe}' awarded to the University of Alabama in Huntsville.


\begin{thebibliography}{9}
\bibitem[Akritas and Bershady 1996]{Akritas1996} Akritas, M.~G., Bershady, M.~A.\ 1996.\ Linear Regression for Astronomical Data with Measurement Errors and Intrinsic Scatter.\ The Astrophysical Journal 470, 706. doi:10.1086/17790110.48550/arXiv.astro-ph/9605002

\bibitem[Arnaud 1996]{arnaud1996} {Arnaud}, K.\ {XSPEC: The First Ten Years}.\ 1996.\  In G.~H. {Jacoby} and J.~{Barnes}, editors, {\em Astr. Data Analysis Software and Systems V}, 101, 17

\bibitem[Baddeley et al. 2015]{Baddeley15}  Baddeley, A., Rubak,E.,  Turner, R. Spatial Point Patterns: Methodology and Applications with R.\ Chapman \& Hall. ISBN: 987-1-4822-1021-7

\bibitem[Bailer-Jones 2017]{bailerjones2017} Bailer-Jones, C.\ 2017.\ {\em Practical Bayesian Inference: A Primer for Physical Scientists}.\ Cambridge University Park.

\bibitem[Bevington 1969]{bevington1969}
{Bevington}, P.\ 1969.\ {\em Data reduction and error analysis for the physical sciences}. McGraw Hill

\bibitem[Bonamente et al. 2004]{bonamente2004} Bonamente, M., Joy, M.~K., Carlstrom, J.~E., Reese, E.~D., LaRoque, S.~J.\ 2004.\ Markov Chain Monte Carlo Joint Analysis of Chandra X-Ray Imaging Spectroscopy and Sunyaev-Zel'dovich Effect Data.\ The Astrophysical Journal 614, 56–63. doi:10.1086/42342010.48550/arXiv.astro-ph/0403016

\bibitem[Bonamente et al. 2006]{bonamente2006} Bonamente, M., Joy, M.~K., LaRoque, S.~J., Carlstrom, J.~E., Reese, E.~D., Dawson, K.~S.\ 2006.\ Determination of the Cosmic Distance Scale from Sunyaev-Zel'dovich Effect and Chandra X-Ray Measurements of High-Redshift Galaxy Clusters.\ The Astrophysical Journal 647, 25–54. doi:10.1086/50529110.48550/arXiv.astro-ph/0512349

\bibitem[Bonamente 2022]{bonamente2022book}
M.~{Bonamente}. {\em Statistics and Analysis of Scientific Data}. Springer, 3rd ed. 

\bibitem[Bonamente \& Spence 2022]{Bonamente22} Bonamente, M., Spence, D. A semi-analytical solution to the maximum-likelihood fit of Poisson data to a linear model using the Cash statistic.\ Journal of Applied Statistics, 49, 522-552. doi:10.1080/02664763.2020.1820960

\bibitem[Box et al. 2015 (63K citations)]{Box15} Box, G., Jenkins, G., Reinsel, G., Ljung, G., {\em Time Series Analysis: Forecasting and Control}, 5th ed., Wiley, 2015. 

\bibitem[Bowyer et al. 1964]{Bowyer64} Bowyer, S., Byram, E., Chubb, T., Friedman, H.\ 1964.\ Lunar Occulation of X-ray Emission from the Crab Nebula.\ Science, 146, 912.\ doi:10.1126/science.146.3646.912

\bibitem[Broos et al. 2010]{Broos10} Broos, P.~S., Townsley, L.~K., Feigelson, E.~D., Getman, K.~V., Bauer, F.~E., Garmire, G.~P.\ 2010.\ Innovations in the Analysis of Chandra-ACIS Observations.\ The Astrophysical Journal 714, 1582–1605. doi:10.1088/0004-637X/714/2/1582

\bibitem[Buchner et al. 2014]{Buchner14} Buchner, J. and 9 colleagues 2014.\ X-ray spectral modelling of the AGN obscuring region in the CDFS: Bayesian model selection and catalogue.\ Astronomy and Astrophysics 564. doi:10.1051/0004-6361/201322971

\bibitem[Cameron and Trivedi 2013 (11K citations)]{cameron2013} Cameron, A. C., Trivedi, P. K. (2013). Regression Analysis of Count Data. United States: Cambridge University Press.

\bibitem[Cash 1976]{cash1976} Cash, W.\ 1976.\  Generation of Confidence Intervals for Model Parameters in X-ray Astronomy.\  Astronomy \& Astrophysics, 52, 307.

\bibitem[Cash 1979]{Cash79} Cash, W.\ 1979.\ Parameter estimation in astronomy through application of the likelihood ratio..\ The Astrophysical Journal 228, 939–947. doi:10.1086/156922

\bibitem[Connors et al. 2011]{Connors11} Connors, A., Stein, N.~M., van Dyk, D., Kashyap, V., Siemiginowska, A.\ 2011.\ LIRA {\textemdash} The Low-Counts Image Restoration and Analysis Package: A Teaching Version via R.\ Astronomical Data Analysis Software and Systems XX 442, 463.

\bibitem[Damiani et al. 1997]{Damiani97} Damiani, F., Maggio, A., Micela, G., Sciortino, S.\ 1997.\ A Method Based on Wavelet Transforms for Source Detection in Photon-counting Detector Images. I. Theory and General Properties.\ The Astrophysical Journal 483, 350–369. doi:10.1086/304217

\bibitem[de Jager et al. 1989]{deJager89} de Jager, O.~C., Raubenheimer, B.~C., Swanepoel, J.~W.~H.\ 1989.\ A powerful test for weak periodic signals with unknown light curve shape in sparse data..\ Astronomy and Astrophysics 221, 180–190.

\bibitem[Diehl and Statler 2006]{Diehl06} Diehl, S., Statler, T.~S.\ 2006.\ Adaptive binning of X-ray data with weighted Voronoi tessellations.\ Monthly Notices of the Royal Astronomical Society 368, 497–510. doi:10.1111/j.1365-2966.2006.10125.x

\bibitem[Eadie et al. 1971]{eadie1971}
W.T. {Eadie}, D.~{Drijiard}, M~{James}, M.~{Roos}, and B.~{Sadoulet}.\ 1971.\  {\em Statistical Methods in Experimental Physics}. North Holland

\bibitem[Ebeling et al. 2006]{Ebeling06} Ebeling, H., White, D.~A., Rangarajan, F.~V.~N.\ 2006.\ ASMOOTH: a simple and efficient algorithm for adaptive kernel smoothing of two-dimensional imaging data.\ Monthly Notices of the Royal Astronomical Society 368, 65–73. doi:10.1111/j.1365-2966.2006.10135.x

\bibitem[Edelson and Krolik 1988]{Edelson88} Edelson, R.~A., Krolik, J.~H.\ 1988.\ The Discrete Correlation Function: A new method for analyzing unevenly sampled variability data.\ The Astrophysical Journal 333, 646. doi:10.1086/166773

\bibitem[En{\ss}lin 2019]{Ensslin2019}  En{\ss}lin, T.~A.\ 2019.\ Information Theory for Fields.\ Annalen der Physik 531, 1800127. doi:10.1002/andp.201800127

\bibitem[Feigelson \& Nelson 1985]{Feigelson85} Feigelson, E.~D., Nelson, P.~I.\ 1985.\ Statistical methods for astronomical data with upper limits. I. Univariate distributions..\ The Astrophysical Journal 293, 192–206. doi:10.1086/163225

\bibitem[Feigelson and Babu 2012]{feigelson2012} {Feigelson}, E. and {Babu}, G.~J.\ 2012.\ {\em Modern Statistical Methods for Astronomy with R Applications}. Cambridge University Press

\bibitem[Feigelson et al. 2022]{Feigelson22} Feigelson, E.~D., Kashyap, V.~L., Siemiginowska, A.\ 2022.\ Time domain methods for X-ray and gamma-ray astronomy.  In Handbook for X-ray and Gamma-Ray Astrophysics, Volume 4: Analysis techniques, Section XVIII: Timing Analysis (Belloni \& Bhattacharya, eds., Springer)

\bibitem[Freeman et al. 2002]{Freeman02} Freeman, P.~E., Kashyap, V., Rosner, R., Lamb, D.~Q.\ 2002.\ A Wavelet-Based Algorithm for the Spatial Analysis of Poisson Data.\ The Astrophysical Journal Supplement Series 138, 185–218. doi:10.1086/324017

\bibitem[Gehrels 1986]{Gehrels1986} Gehrels, N.\ 1986.\ Confidence Limits for Small Numbers of Events in Astrophysical Data.\ The Astrophysical Journal 303, 336. doi:10.1086/164079

\bibitem[Giacconi et al. 1962]{Giacconi1962} Giacconi, R., Gursky, H., Paolini, F., Rossi, B.\ 1962.\ Evidence for X-rays from Sources Outside the Solar System.\ Physical Review Letters, 9, 439. doi:10.1103/PhysRevLett.9.439

\bibitem[Hearn 1968]{Hearn68}  Hearn, D.\ 1968.\ A Search for Celestial Sources of Gamma Rays of Energy Greater than 100 MeV.\ Smithsonian Astrophysical Observatory Special Report 277.

\bibitem[Hilbe et al. 2017]{hilbe2017} Hilbe, J., de Souza, R., Ishida, E.\ 2017.\ {\em Bayesian Models for Astrophysical Data: Using R, JAGS, Python, and Stan}. Cambridge University Press

\bibitem[Hogg, Tanis and Zimmerman 2023]{hogg2023}
R.~{Hogg}, E.~{Tanis}, and D.~{Zimmerman}.\ 2023.\ {\em Probability and Statistical Inference}. Pearson, Tenth Edition.

\bibitem[Isobe et al. 1986]{Isobe86} Isobe, T., Feigelson, E.~D., Nelson, P.~I.\ 1986.\ Statistical Methods for Astronomical Data with Upper Limits. II. Correlation and Regression.\ The Astrophysical Journal 306, 490. doi:10.1086/164359

\bibitem[Ivezić et al. 2019]{ivezic2019}  Ivezić, Z., Connolly, A., VanderPlas, J., Gray, A.\ 2019.\ {\em Statistics, Data Mining, and Machine Learning in Astronomy: A Practical Python Guide for the Analysis of Survey Data}, Princeton University Press 

\bibitem[Kaastra 1996]{kaastra1996} J.~S. {Kaastra}, R.~{Mewe}, and H.~{Nieuwenhuijzen}\ 1996.\ {SPEX: a new code for spectral analysis of X and UV spectra.} In K.~{Yamashita} and T.~{Watanabe}, editors, {\em UV and X-ray Spectroscopy of Astrophysical and Laboratory Plasmas}, pages 411--414
  
\bibitem[Kashyap et al. 2010]{Kashyap10} Kashyap, V.~L. and 6 colleagues 2010.\ On Computing Upper Limits to Source Intensities.\ The Astrophysical Journal 719, 900–914. doi:10.1088/0004-637X/719/1/900

\bibitem[Kelly 2007]{Kelly07} Kelly, B.~C.\ 2007.\ Some Aspects of Measurement Error in Linear Regression of Astronomical Data.\ The Astrophysical Journal 665, 1489–1506. doi:10.1086/519947

\bibitem[Kraft et al. 1991]{Kraft91} Kraft, R.~P., Burrows, D.~N., Nousek, J.~A.\ 1991.\ Determination of Confidence Limits for Experiments with Low Numbers of Counts.\ The Astrophysical Journal 374, 344. doi:10.1086/170124

\bibitem[Leahy et al. 1983]{Leahy83} Leahy, D.~A., Elsner, R.~F., Weisskopf, M.~C.\ 1983.\ On searches for periodic pulsed emission - The Rayleigh test compared to epoch folding.\ The Astrophysical Journal 272, 256–258. doi:10.1086/161288

\bibitem[Li and Ma 1983]{Li83} Li, T.-P., Ma, Y.-Q.\ 1983.\ Analysis methods for results in gamma-ray astronomy..\ The Astrophysical Journal 272, 317–324. doi:10.1086/161295

\bibitem[Mantz et al.2010]{Mantz2010} Mantz, A., Allen, S.~W., Rapetti, D., Ebeling, H.\ 2010.\ The observed growth of massive galaxy clusters - I. Statistical methods and cosmological constraints.\ Monthly Notices of the Royal Astronomical Society 406, 1759–1772. doi:10.1111/j.1365-2966.2010.16992.x10.48550/arXiv.0909.3098

\bibitem[Mantz et al.2010b]{Mantz2010b} Mantz, A., Allen, S.~W., Ebeling, H., Rapetti, D., Drlica-Wagner, A.\ 2010.\ The observed growth of massive galaxy clusters - II. X-ray scaling relations.\ Monthly Notices of the Royal Astronomical Society 406, 1773–1795. doi:10.1111/j.1365-2966.2010.16993.x10.48550/arXiv.0909.3099


\bibitem[Mattox et al. 1996]{Mattox96} Mattox, J.~R. and 22 colleagues 1996.\ The Likelihood Analysis of EGRET Data.\ The Astrophysical Journal 461, 396. doi:10.1086/177068

\bibitem[Mukherjee et al. 1998]{Mukherjee98} Mukherjee, S., Feigelson, E.~D., Jogesh Babu, G., Murtagh, F., Fraley, C., Raftery, A.\ 1998.\ Three Types of Gamma-Ray Bursts.\ The Astrophysical Journal 508, 314–327. doi:10.1086/306386

\bibitem[Park et al. 2006]{Park06} Park, T. and 6 colleagues 2006.\ Bayesian Estimation of Hardness Ratios: Modeling and Computations.\ The Astrophysical Journal 652, 610–628. doi:10.1086/507406

\bibitem[Percival et al. 2022 ]{Percival22} Percival, W.~J., Friedrich, O., Sellentin, E., Heavens, A.\ 2022.\ Matching Bayesian and frequentist coverage probabilities when using an approximate data covariance matrix.\ Monthly Notices of the Royal Astronomical Society 510, 3207–3221. doi:10.1093/mnras/stab3540

\bibitem[Protassov et al. 2002]{Protassov02} Protassov, R., van Dyk, D.~A., Connors, A., Kashyap, V.~L., Siemiginowska, A.\ 2002.\ Statistics, Handle with Care: Detecting Multiple Model Components with the Likelihood Ratio Test.\ The Astrophysical Journal 571, 545–559. doi:10.1086/339856

\bibitem[Rossi 1948]{Rossi48} Rossi, B.\ 1948.\  Interpretation of Cosmic-Ray Phenomena, Reviews Modern Physics, 20, 537. doi:10.1103/RevModPhys.20.537

\bibitem[Scargle 1982]{Scargle82} Scargle, J.~D.\ 1982.\ Studies in astronomical time series analysis. II. Statistical aspects of spectral analysis of unevenly spaced data..\ The Astrophysical Journal 263, 835–853. doi:10.1086/160554

\bibitem[Scargle 1998]{Scargle98} Scargle, J.~D.\ 1998.\ Studies in Astronomical Time Series Analysis. V. Bayesian Blocks, a New Method to Analyze Structure in Photon Counting Data.\ The Astrophysical Journal 504, 405–418. doi:10.1086/306064

\bibitem[Scargle et al. 2013]{Scargle13} Scargle, J.~D., Norris, J.~P., Jackson, B., Chiang, J.\ 2013.\ Studies in Astronomical Time Series Analysis. VI. Bayesian Block Representations.\ The Astrophysical Journal 764. doi:10.1088/0004-637X/764/2/167

\bibitem[Schmitt 1985]{Schmitt85} Schmitt, J.~H.~M.~M.\ 1985.\ Statistical analysis of astronomical data containing upper bounds: general methods and examples drawn from X-ray astronomy..\ The Astrophysical Journal 293, 178–191. doi:10.1086/163224

\bibitem[Selig et al. 2015]{Selig15} Selig, M., Vacca, V., Oppermann, N., En{\ss}lin, T.~A.\ 2015.\ The denoised, deconvolved, and decomposed Fermi {\ensuremath{\gamma}}-ray sky. An application of the D$^{3}$PO algorithm.\ Astronomy and Astrophysics 581. doi:10.1051/0004-6361/201425172

\bibitem[Stein et al. 2015]{Stein15} Stein, N.~M., van Dyk, D.~A., Kashyap, V.~L., Siemiginowska, A.\ 2015.\ Detecting Unspecified Structure in Low-count Images.\ The Astrophysical Journal 813. doi:10.1088/0004-637X/813/1/66

\bibitem[Tousey et al. 1951]{Tousey51} Tousey, R., Watanabe, K., Purcell, J. D.\ 1951.\ Measurements of Solar Extreme Ultraviolet and X-Rays from Rockets by Means of a CoSO4:Mn Phosphor.\ Physical Review, 83, 792-797. doi: 
10.1103/PhysRev.83.792 

\bibitem[Uttley et al. 2005]{Uttley05} Uttley, P., McHardy, I.~M., Vaughan, S.\ 2005.\ Non-linear X-ray variability in X-ray binaries and active galaxies.\ Monthly Notices of the Royal Astronomical Society 359, 345–362. doi:10.1111/j.1365-2966.2005.08886.x

\bibitem[VanderPlas 2018]{VanderPlas18} VanderPlas, J.~T.\ 2018.\ Understanding the Lomb-Scargle Periodogram.\ The Astrophysical Journal Supplement Series 236. doi:10.3847/1538-4365/aab766

\bibitem[van Dyk et al. 2001]{vanDyk01} van Dyk, D.~A., Connors, A., Kashyap, V.~L., Siemiginowska, A.\ 2001.\ Analysis of Energy Spectra with Low Photon Counts via Bayesian Posterior Simulation.\ The Astrophysical Journal 548, 224–243. doi:10.1086/318656

\bibitem[Vaughan et al. 2003]{Vaughan03} Vaughan, S., Edelson, R., Warwick, R.~S., Uttley, P.\ 2003.\ On characterizing the variability properties of X-ray light curves from active galaxies.\ Monthly Notices of the Royal Astronomical Society 345, 1271–1284. doi:10.1046/j.1365-2966.2003.07042.x

\bibitem[Vaughan 2005]{Vaughan05} Vaughan, S.\ 2005.\ A simple test for periodic signals in red noise.\ Astronomy and Astrophysics 431, 391–403. doi:10.1051/0004-6361:20041453

\bibitem[Wall and Jenkins 2012]{wall2012}
J.~V. {Wall} and C.~R. {Jenkins}.
\newblock {\em Practical Statistics for Astronomers}.
\newblock Cambridge, Second Edition, 2012.

\bibitem[Xu et al. 2014]{Xu14} Xu, J. and 8 colleagues 2014.\ A Fully Bayesian Method for Jointly Fitting Instrumental Calibration and X-Ray Spectral Models.\ The Astrophysical Journal 794. doi:10.1088/0004-637X/794/2/97

\bibitem[Xu et al. 2021]{Xu21} Xu, C., G{\"u}nther, H.~M., Kashyap, V.~L., Lee, T.~C.~M., Zezas, A.\ 2021.\ Change-point Detection and Image Segmentation for Time Series of Astrophysical Images.\ The Astronomical Journal 161. doi:10.3847/1538-3881/abe0b6

\bibitem[Zhang et al. 2008]{Zhang08} Zhang, B., Fadili, J.~M., Starck, J.-L.\ 2008.\ Wavelets, Ridgelets, and Curvelets for Poisson Noise Removal.\ IEEE Transactions on Image Processing 17, 1093–1108. doi:10.1109/TIP.2008.924386

\end{thebibliography}
\end{document}